\documentclass[runningheads,twoside]{llncs}

\usepackage[ansinew]{inputenc}
\usepackage[T1]{fontenc}

\usepackage{graphicx}
\usepackage{amssymb}
\usepackage{textcomp}
\usepackage{amscd}
\usepackage{amsmath}
\usepackage[usenames]{color}
\definecolor{sonnengelb}{rgb}{0.9,0.5,0} 
\definecolor{brown}{rgb}{0.5,1,0}
\usepackage{algorithm,algorithmic}
\usepackage[ruled,vlined,linesnumbered,algo2e]{algorithm2e}

\begin{document}
\mainmatter              
\title{How to Attack the NP-complete Dag Realization Problem in Practice\thanks{This work
    was supported by the DFG Focus Program Algorithm Engineering,
    grant MU 1482/4-2. Extended abstract is to appear in Proceedings of SEA 
2012, LNCS, Springer.}}
\titlerunning{How to Attack the Dag Realization Problem}
\author{Annabell Berger and Matthias M\"uller-Hannemann}
\institute{
  Dept.\ of Computer Science,
  Martin-Luther-Universit\"at Halle-Wittenberg\\
\email{\{berger,muellerh\}@informatik.uni-halle.de}}
\authorrunning{A. Berger, M. M\"uller-Hannemann}
\maketitle

\begin{abstract}
We study the following fundamental realization problem 
of directed acyclic graphs (dags). 
Given a sequence 
 $S:={a_1 \choose b_1},\dots,{a_n \choose b_n}$
with $a_i,b_i\in \mathbb{Z}_0^+$, does there exist a dag (no parallel arcs allowed) with 
labeled vertex set $V:=\{v_1,\dots,v_n\}$ such that for all $v_i\in V$
indegree and outdegree of $v_i$ match exactly 
the given numbers $a_i$ and $b_i$,
respectively?
Recently this decision problem has been shown to be NP-complete 
by Nichterlein~\cite{Nichterlein:2011}. However, we can show that several important 
classes of sequences are efficiently solvable.
In previous work~\cite{BergerMueller-Hannemann-FCT2011}, we have proved that yes-instances always have a special kind of topological order which allows us to reduce the number of possible topological orderings in most cases drastically. This leads to an exact 
exponential-time algorithm
which significantly improves upon a straightforward approach.
Moreover, a combination of this exponential-time algorithm with a special strategy gives a linear-time algorithm.
Interestingly, in systematic experiments we observed that we could solve a huge majority of all instances by the linear-time heuristic. This motivates us to develop characteristics 
like dag density and ``distance to provably easy sequences'' 
which can give us an indicator how easy or difficult a given sequence can be realized.

Furthermore, we propose a randomized algorithm which exploits our structural insight on topological sortings and uses a number of reduction rules.
We compare this algorithm with other straightforward 
randomized algorithms in extensive experiments. We 
observe that it clearly outperforms all other variants 
and behaves surprisingly well for almost all instances.
Another striking observation is that our simple linear-time algorithm 
solves a set of real-world instances from different domains, namely ordered binary decision diagrams (OBDDs), train and flight schedules, as well as instances derived from food-web networks without any exception.
\end{abstract}


\section{The Dag Realization Problem}\label{Introduction}




\vspace*{-1ex}

\noindent
{\bf Dag realization problem:}
Given is a finite sequence $S:={a_1 \choose b_1},\dots,{a_n \choose b_n}$
with $a_i,b_i\in \mathbb{Z}_0^+.$ Does there exist an acyclic digraph
(without parallel arcs) $G=(V,A)$ with 
the labeled vertex set $V:=\{v_1,\dots,v_n\}$ such that we have 
indegree $d^-_G(v_i) = a_i$ and outdegree $d^+_G(v_i) = b_i$ 
for all $v_i\in V$?\\

If the answer is ``yes'', we call sequence $S$ \emph{dag sequence} and
the acyclic digraph $G$ (a so-called ``dag'') a \emph{dag
  realization}. A relaxation of this problem -- not demanding the acyclicity of digraph $G$ -- is called \emph{digraph realization problem}. In this case, we call  $G$ \emph{digraph realization} and $S$  \emph{digraph sequence}. The digraph realization problem can be solved in linear-time using an algorithm by Wang and Kleitman \cite{KleitWang:73}. Unless explicitly stated, we assume that a sequence does not contain any \emph{zero tuples} ${0 \choose 0}$. Moreover, we will tacitly assume that $\sum_{i=1}^{n}a_i=\sum_{i=1}^{n}b_i$, as this is obviously a necessary condition for any realization to exist, since the number of ingoing arcs must equal the number of outgoing arcs. Furthermore, we denote tuples ${a_i \choose b_i}$ with $a_i>0$ and $b_i=0$ as \emph{sink tuples}, those with $a_i=0$ and $b_i>0$ as \emph{source tuples}, and the remaining ones with $a_i>0$ and $b_i>0$ as \emph{stream tuples}. We call a sequence only consisting of source and sink tuples, \emph{source-sink-sequence}. A sequence $S={a_1 \choose b_1},\dots,{a_n \choose b_n}$ with $q$ source tuples and $s$ sink tuples is denoted as \emph{canonically sorted}, if and
only if the first $q$ tuples in this labeling are decreasingly sorted source tuples (with respect to the $b_i$) and the last $s$ tuples are increasingly sorted sink tuples (with respect to the $a_i$).\\[-1ex] 

\noindent
{\bf Hardness and efficiently solvable special cases.}
Nichterlein very recently showed that the dag realization problem is NP-complete \cite{Nichterlein:2011}. On the other hand, there are several classes of sequences for which the problem is not hard. One of these sequences are source-sink-sequences, for which one only has 
to find a digraph realization.
The latter is already a dag realization, since no vertex has incoming as well as outgoing arcs.  Furthermore, sparse sequences with $\sum_{i=1}^{n}a_i\leq n-1$ are polynomial-time solvable as we will show below. We denote such sequences by \emph{forest sequences}.
The main difficulty for the dag realization problem is to find out a ``topological ordering of the sequence''. In the case where we have one, our problem is nothing else but a directed $f$-factor problem on a complete dag. The labeled vertices of this complete dag are ordered in the given topological order. This problem can be reduced to a bipartite undirected $f$-factor problem which can be solved in polynomial time via a further famous reduction by Tutte \cite{Tutte52} to a bipartite perfect matching problem. In a previous paper \cite{BergerMueller-Hannemann-FCT2011}, we  proved that a certain ordering of a special class of sequences --\emph{opposed sequences}-- always leads to a topological ordering of the tuples for at least one dag realization of a given dag sequence. On the other hand, it is not necessary to apply the reduction via Tutte if we possess one possible topological ordering of a dag sequence. The solution is much easier. Next, we describe our approach.\\[-1ex]

\noindent
{\bf Realization with a prescribed topological order.}
We denote a dag sequence $S:={a_1 \choose b_1},\dots,{a_n \choose b_n}$ which possesses a dag realization with a topological numbering corresponding to the increasing numbering of its tuples by \emph{dag sequence for a given topological order} and analogously the digraph $G=(V,A)$ by \emph{dag realization for a given topological order}. Without loss of generality, we may assume that the source tuples come first in the prescribed numbering and are ordered decreasingly with respect to their $b_i$ values.
A realization algorithm works as follows. Consider the first tuple ${a_{q+1} \choose b_{q+1}}$ from the prescribed topological order  which is not a source tuple. Then there must exist $a_{q+1}$ source tuples with a smaller number in the given dag sequence. Reduce the $a_{q+1}$ first (i.e.\ with largest $b_i$) source tuples by one and set the indegree of tuple ${a_{q+1} \choose b_{q+1}}$ to $0.$ That means, we reduce sequence $S:={a_1 \choose b_1},\dots,{a_{q+1} \choose b_{q+1}},\dots,{a_n \choose b_n}$ to sequence $S':={a_1 \choose b_1-1},\dots,{a_{a_{q+1}} \choose b_{a_{q+1}}-1},\dots,{a_q \choose b_q},{0 \choose b_{q+1}},\dots,{a_n \choose b_n}.$ If we get zero tuples in $S',$ then we delete them and denote the new sequence for simplicity also by $S'.$ Furthermore, we label this sequence with a new numbering starting from one to its length and consider this sorting as the given topological ordering for $S'$. We repeat this process until we get an empty sequence (corresponding to the realizability of $S$) or get stuck (corresponding to the non-realizability of $S$). The correctness of our algorithm is proven in Lemma~\ref{TH:topologicalRealization}.  

\begin{lemma}\label{TH:topologicalRealization}
$S$ is a dag sequence for a given topological order $\Leftrightarrow$ $S'$ is a dag sequence for its corresponding topological order.
\end{lemma}

\noindent
{\bf Discussion of our main theorem and its corresponding algorithm.}
We do not know how to determine a feasible topological ordering (i.e., one corresponding to a realization) for an arbitrary dag sequence. However, we are able to restrict the types of possible permutations of the tuples. For that, we need the following order relation $\leq_{opp}\subset \mathbb{Z}^2\times \mathbb{Z}^2$, introduced in \cite{BergerMueller-Hannemann-FCT2011}. 

\begin{definition}[opposed relation]
Given are $c_1:={a_1 \choose b_1}\in \mathbb{Z}^2$ and $c_2:={a_2 \choose b_2}\in \mathbb{Z}^2.$ We define: $c_1\leq_{opp}c_2 \Leftrightarrow (a_1 \leq a_2 \land b_1\geq b_2).$
\end{definition}

Note, that a pair $c_1$ equals $c_2$ with respect to the opposed relation if and only if $a_1=a_2$ and $b_1=b_2.$
The opposed relation is reflexive, transitive and antisymmetric and therefore a partial, but not a total order. 
Our following theorem leads to a recursive algorithm with exponential running time and results in Corollary~\ref{KorollarTopologischeRealisation} which proves the existence of a special type of possible topological sortings provided that sequence $S$ is a dag sequence. 

\begin{theorem}[main theorem \cite{BergerMueller-Hannemann-FCT2011}]\label{RealisationDagSequenzen}
Let $S$ be a canonically sorted sequence containing $k>0$ source tuples. Furthermore, we assume that $S$ is not a source-sink-sequence. We define the set 
$$V_{min}:=\left\{{a_i\choose b_i}|~{a_i \choose b_i}
\begin{array}{l}
\textnormal{ is stream tuple}, a_i\leq k, \\ 
\textnormal{ and there is no stream tuple }
\end{array}
{a_j\choose b_j} <_{opp} {a_{i} \choose b_{i}}\right\}.$$
$S$ is a dag sequence if and only if $V_{min}\neq \emptyset$ and there exists an element ${a_{\ell}\choose b_{\ell}}\in V_{min}$ such that $S':=$
$${0 \choose b_1{-}1},\dots,{0 \choose b_{a_{\ell}}{-}1},{0 \choose b_{a_{\ell}+1}},\dots,{0 \choose b_k},\dots,{a_{\ell-1} \choose b_{\ell-1}},{0 \choose b_{\ell}},{a_{\ell+1} \choose b_{\ell+1}},\dots,{a_n \choose b_n}$$
is a dag sequence.
\end{theorem}

\begin{algorithm2e}
\SetKwInOut{Input}{Input}
\SetKwInOut{Output}{Output}
\Input{A canonically sorted sequence $S.$}
\Output{A Boolean flag indicating whether $S$ is realizable.}
\eIf{$S$ is not a source-sink-sequence}{
  count the number of sources in $S$ and determine set $V'_{min}$\;
  \For{all ${a_j \choose b_j} \in V'_{min}$}{
    create a working copy $S'$ of $S$ with tuples ${a'_i \choose b'_i} = {a_i \choose b_i}$\;
    set $b'_i \leftarrow b'_i-1$ for $a'_j$ largest sources ${0 \choose b'_i}$\;
    set $a'_j \leftarrow 0$\;
    delete ${0 \choose 0}$-tuples\;
    \lIf{DagRealization($S'$)}{return TRUE}\;
  }
  return FALSE\;
}(\tcp*[f]{Realization of a source-sink-sequence})
{
  \While{the set of source tuples in $S$ is not empty}{
    choose a largest source tuple ${0 \choose b_j}$\;
    \lIf{number of sinks in $S$ is smaller than $b_j$}{return FALSE}\;
    set $a_i \leftarrow a_i-1$ for $b_j$ largest sinks ${a_i \choose 0}$\;
    delete ${0 \choose 0}$-tuples\;
  }
  return TRUE\;
}

\caption{DagRealization(sequence $S$)} \label{alg:dag realization}
\end{algorithm2e}

                

Sequence $S'$ may contain zero tuples.
If this is the case, we delete them and call the new sequence for simplicity also $S'$.
Theorem \ref{RealisationDagSequenzen} ensures the possibility for reducing a dag sequence into a source-sink-sequence.
The latter can be realized by using the algorithm for realizing digraph sequences \cite{KleitWang:73}. The whole algorithm is summarized in Algorithm~\ref{alg:dag realization}, where we consider the maximum subset $V'_{min}$ of $V_{min}$ only containing pairwise disjoint stream tuples. 
The bottleneck of this approach is the size of set $V'_{min}.$
We give an example for the execution of Algorithm~\ref{alg:dag realization} 
in the Appendix.
Our pseudocode does not specify 
the order in which we process the elements of $V'_{min}$ in line 3. 
Several strategies are possible which have a significant influence on the overall performance.
The most promising deterministic strategy (as we will learn in the next sections) is to use the lexicographic order, starting with the lexicographic maximum element within $V'_{min}$. 
 In \cite{BergerMueller-Hannemann-FCT2011} we introduced a special class of dag sequences -- \emph{opposed sequences} -- where we have $|V'_{min}|=1,$ if sequence $S$ is not a source-sink-sequence. We call a sequence $S$  \emph{opposed sequence}, if it is possible to sort its stream tuples in such a way, that $a_i\leq a_{i+1}$ and $b_i\geq b_{i+1}$ is valid for stream tuples with indices $i$ and $i+1.$ In this case, we have the property ${a_i \choose b_i}\leq_{opp}{a_{i+1}\choose b_{i+1}}$ for all stream tuples. At the beginning of the sequence we insert all source tuples such that the $b_i$ build a decreasing sequence and at the end of sequence $S$ we put all sink tuples in increasing ordering with respect to the corresponding $a_i.$ The notion \textit{opposed  sequence} describes a sequence, where it is possible to compare all stream tuples among each other and to put them in a ``chain''. Indeed, this is not always possible because the opposed order is not a total order. However, for opposed sequences line (3) to line (9) in Algorithm~\ref{alg:dag realization} are 
executed at most once in each recursive call, because we have always $|V'_{min}| \leq 1.$ Overall, we obtain a linear-time algorithm for opposed sequences. However, there are many sequences which are not opposed, but Theorem~\ref{RealisationDagSequenzen} still yields a polynomial decision time. Consider for example dag sequence $S:={0  \choose 3},{0 \choose 3},{2 \choose 2},{3 \choose 3},{1 \choose 0},{2 \choose 0},{3 \choose 0}$ which is not an opposed sequence, because stream tuples ${2 \choose 2}$ and ${3 \choose 3}$ are not comparable with respect to the opposed ordering. However, we have $|V'_{min}|=|\{{2 \choose 2}\}|=1$ and so we reduce $S$ to $S'={0
  \choose 2},{0 \choose 2},{0 \choose 2},{3 \choose 3},{1 \choose   0},{2 \choose 0},{3 \choose 0}$, leading to the realizable source-sink-sequence ${0 \choose 1},{0 \choose 1},{0 \choose 1},{0   \choose 3},{1 \choose 0},{2 \choose 0},{3 \choose 0}.$ 
Theorem \ref{RealisationDagSequenzen} leads to further interesting insights.
We can prove the existence of special topological sortings.
 
 \begin{corollary}[\cite{BergerMueller-Hannemann-FCT2011}]\label{KorollarTopologischeRealisation}
For every dag sequence $S$, there exists a dag realization
$G=(V,A)$ with a topological ordering $v_{l_1}, \dots, v_{l_{n_s}}$
of all $n_s$ vertices corresponding to stream tuples, such that we
cannot find ${a_{l_j} \choose b_{l_j}}<_{opp}{a_{l_i} \choose
  b_{l_i}}$ for $l_i < l_j.$ 
\end{corollary}

We call a topological ordering of a dag sequence obeying the conditions in Corollary~\ref{KorollarTopologischeRealisation} an \emph{opposed topological sorting}.
At the beginning of our work (when the complexity of the dag realization problem was still 
open), we conjectured that the choice of the lexicographical largest tuple from $V'_{min}$ in line (3) would solve our problem in polynomial time. We call this approach \emph{lexmax strategy} and a dag sequence which is realizable with this strategy \emph{lexmax sequence}, otherwise we call 
it \emph{non-lexmax sequence}. Hence, we conjectured  the following.

\begin{conjecture}[lexmax conjecture]\label{Co:lexmaxconjecture}
Each dag sequence is a lexmax sequence.
\end{conjecture}

We soon 
disproved our own conjecture by a counter-example 
(Example~\ref{example}, described in the following 
section and in Appendix~\ref{app:examples}).
In systematic experiments we found out that a large fraction of sequences can be solved by this strategy in polynomial time. 
We tell this story in the next Section~\ref{StoryExperiments}. Moreover, we use the structural insights from our main theorem 
to develop
a randomized algorithm which performs well in practice (Section~\ref{RandomisierteAlgorithmen}). 
Proofs and further supporting material can be found in the Appendix and 
in \cite{BergerPhD2011}. 

\section{Lessons from Experiments with the Lexmax Strategy}
\label{StoryExperiments}



\noindent
{\bf Why we became curious.}
To see whether our lexmax Conjecture~\ref{Co:lexmaxconjecture} might be true, we generated a set of dag sequences, called \emph{randomly generated sequences} in the sequel, by the following principle: Starting with a complete acyclic digraph, delete $k$ of its arcs uniformly at random. We take the degree sequence from the resulting graph. Note that we only sample uniformly with respect to random dags but not uniformly degree sequences since degree sequences have different numbers of corresponding dag realizations.  
In a first experiment we created with the described process one million dag sequences with 20 tuples each, and $m= \sum_{i=1}^{20}a_i=114$. Likewise, we built up  another million dag sequences with 25 tuples and $\sum_{i=1}^{25}a_i=180$. The fact that the lexmax strategy realized all these test instances without a single failure was quite encouraging. The lexmax conjecture \ref{Co:lexmaxconjecture} seemed to be true, only a correctness proof was missing. But quite soon, in an attempt to prove the conjecture, we artificially constructed a first counter-example, a dag sequence which is definitely no lexmax sequence, as can easily be verified:
\begin{example}\label{example}
$S := {0 \choose 3},{0 \choose 1},{1 \choose 2},{2 \choose 3},{4 \choose 4},{1 \choose 1},{1 \choose 0},{2 \choose 0},{3 \choose 0}$. 
Details are shown in Appendix~\ref{app:examples}.
The rightmost path in the recursion tree shown in Figure~\ref{fig:example}
corresponds to the lexmax strategy, but is unsuccessful. 
\end{example}

Even worse: we also found an example (Example~\ref{Ex:StrategieSichtweise}) showing that no fixed strategy which chooses an element from $V'_{min}$ in Algorithm \ref{alg:dag realization} and does not consider the corresponding set of sinks, will fail in general.
 
 \begin{example}\label{Ex:StrategieSichtweise}
 We consider the two sequences  
 $$S_1:={0\choose 5},{0 \choose 5},{0 \choose 5},{0 \choose 2},{0 \choose 2},{5 \choose 5},{5 \choose 5},{2 \choose 2},{2 \choose 2},{1 \choose 0},{1 \choose 0},{2 \choose 0},{6 \choose 0},{9 \choose 0}$$ and 
 $$S_2:={0\choose 5},{0 \choose 5},{0 \choose 5},{0 \choose 2},{0 \choose 2},{5 \choose 5},{5 \choose 5},{2 \choose 2},{2 \choose 2},{6 \choose 0},{6 \choose 0},{7 \choose 0},$$
 only differing in their sink tuples. Sequence $S_2$ can only be realized by the lexmax strategy,
 while several strategies but not the lexmax strategy work
 for $S_1$. Thus, there is no strategy which can be applied in both cases.
 \end{example}

\begin{figure}[t]
\begin{minipage}{.45\textwidth}
\centering
\includegraphics[width=1.1\textwidth]{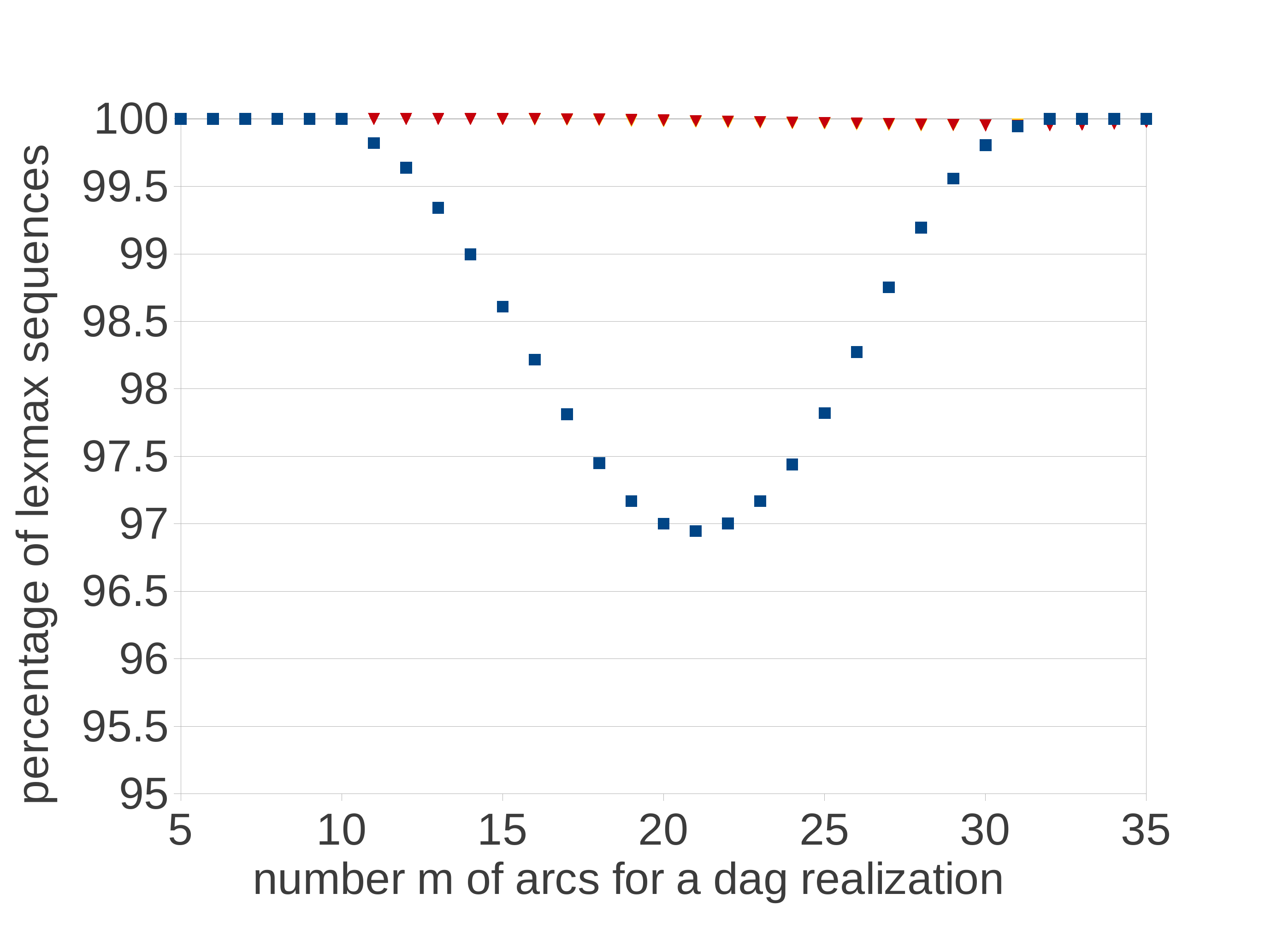}
\caption{\label{fig:PercentageLexMaxSequences} Percentage of (non-trivial) lexmax sequences  
for systematically generated (blue squares) and randomly generated sequences (red triangles) with $9$ tuples and $m \in \{5,\dots,35\}$.}
\end{minipage}
\hfill
\begin{minipage}{.5\textwidth}
\hspace*{-3ex}
\includegraphics[width=1.2\textwidth]{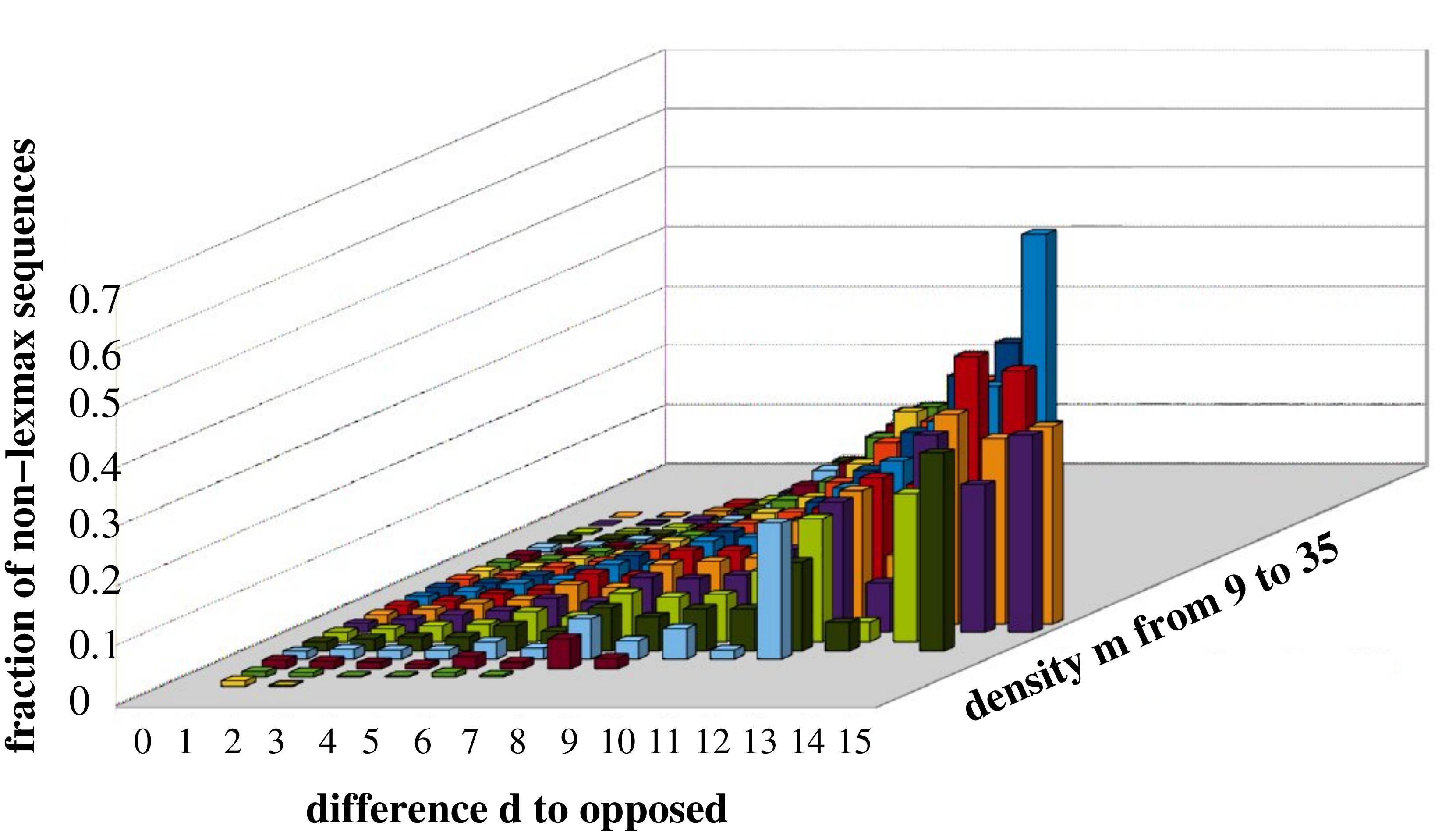}
\caption{\label{fig:OpposedDistanceNonLexMax} Fraction of systematic non-lexmax sequences with $9$ tuples, $m \in \{9,\dots,35\}$, and varying difference to opposed $d(S)$.}
\end{minipage}
\end{figure}

These observations give rise to several immediate questions:
Why did we construct by our sampling method (for $n=20$ and $n=25$) only dag sequences which are lexmax sequences?
How many dag sequences are not lexmax sequences?
Therefore, we started with systematic experiments.
For small instances with $n \in \{7,8,9\}$ tuples we generated systematically the set of all dag sequences
with all possible $\sum_{i=1}^m a_i =: m$, see for an example the case $n=9$ in Figure~\ref{fig:PercentageLexMaxSequences} and Appendix~\ref{app:material}. 
More precisely, we considered only \emph{non-trivial sequences},
i.e.\ we eliminated all source-sink sequences and all sequences with only one stream tuple.
We denote this set by \emph{systematically generated sequences}.  
Note that the number of sequences grows so fast in $n$ that a systematic construction of all sequences 
with a larger size is impossible. 
We observed the following:
\begin{enumerate}
\item 
The fraction of lexmax sequences among the systematically generated sequences is quite high.
For all $m$ it is above $96.5\%$, see Figure~\ref{fig:PercentageLexMaxSequences} (blue squares).
\item 
The fraction of lexmax sequences strongly depends on $m$. It is largest for sparse and dense dags.

\item Lexmax sequences are overrepresented among one million randomly generated sequences (for each $m$), we observe more than $99\%$ for all densities of dags, see Figure~\ref{fig:PercentageLexMaxSequences} (red triangles).
\end{enumerate}

This leads to the following questions: Given a sequence for which we seek a dag realization.
How should we proceed in practice? As we have seen, the huge majority of dag sequences are lexmax sequences. Is it possible to find characteristic properties for lexmax sequences or non-lexmax sequences, respectively?\\[-1ex]

\noindent
{\bf Distance to opposed sequences.}
Let us exploit our characterization  
that opposed sequences are efficiently solvable. We propose the \emph{distance to opposed} $d(S)$ for each dag sequence $S.$ Consider for that the topological order of a dag realization $G$ given by Algorithm~\ref{alg:dag realization}, if in line (3) elements are chosen in decreasing lexicographical order. 
This ordering corresponds to exactly one path of the recursion tree.
Thus, we obtain one unique dag realization $G$ for $S$, if existing.
Now, we renumber dag sequence $S$ such that it follows the topological order
induced by the execution by this algorithm, i.e.\ by the sequence of choices
of elements from $V'_{min}.$
Then the distance to opposed is defined as the number of pairwise incomparable stream tuples with respect to  this order, more precisely, 
$$d(S):=\left|\left\{\left({a_i \choose b_i},{a_j \choose b_j}\right)|~{a_i \choose b_i},{a_j \choose b_j} \begin{array}{l}
\textnormal{ incomparable stream tuples }\\ 
\textnormal{ w.r.t.\ $\leq_{opp}$ and } i<j 
\end{array}
\right\}\right|.$$

\paragraph{Question 1: Do randomly generated sequences possess a preference to a ``small'' distance to opposed in comparison with systematically generated sequences?}

\begin{figure}[t]
\begin{minipage}{.47\textwidth}
\hspace*{-1ex}
\includegraphics[width=1.15\textwidth]{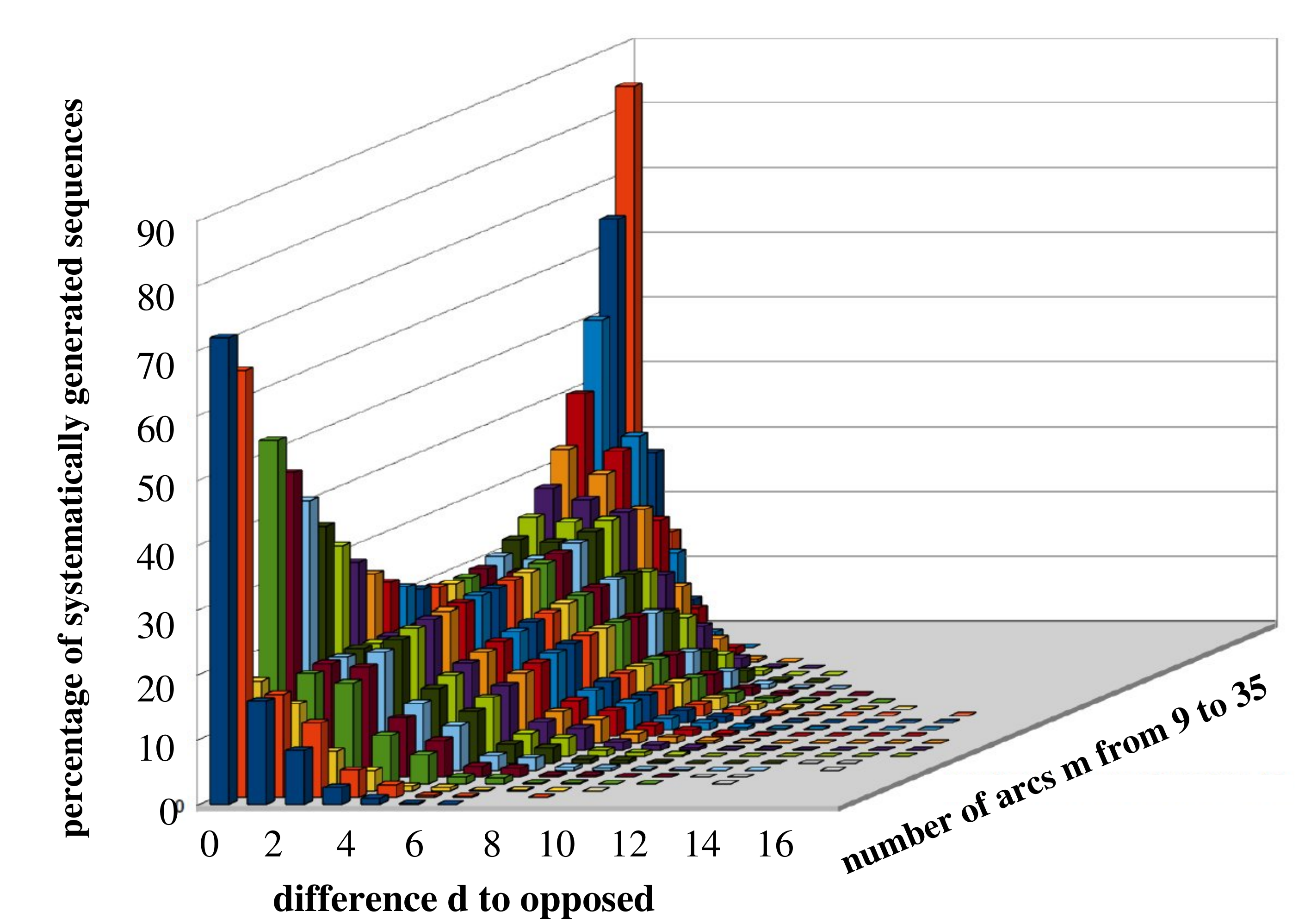}
\end{minipage}
\hfill
\begin{minipage}{.47\textwidth}
\includegraphics[width=1.15\textwidth]{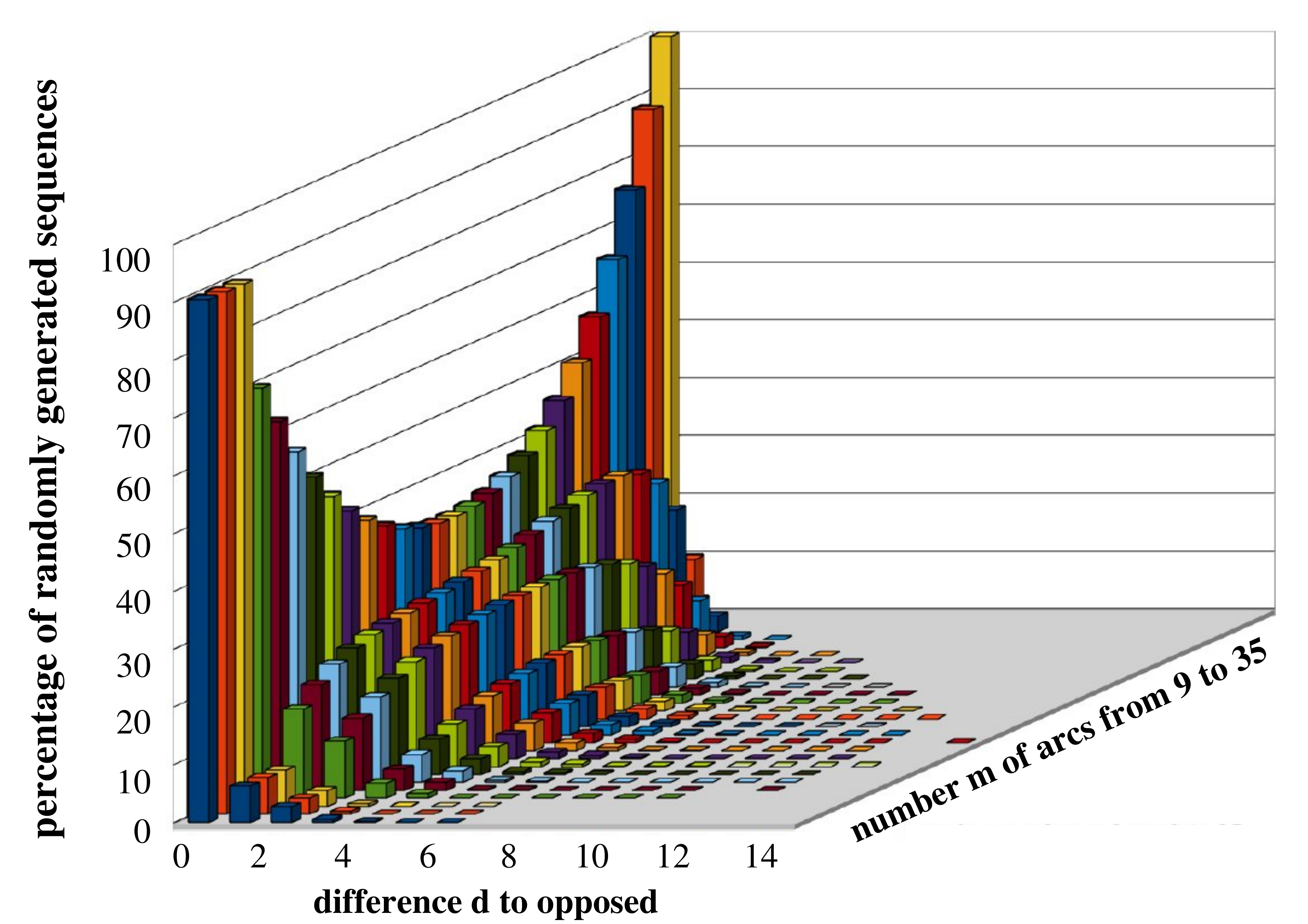}
\end{minipage}
\caption{\label{fig:SystematicDifferenceToOpposed} Percentage of systematically generated sequences $S$ (left) and randomized generated sequences (right)
with their difference $d(S)$ to opposed for $n=9$ tuples and $m \in \{9,\dots,35\}$.}

\end{figure}

In Figure~\ref{fig:SystematicDifferenceToOpposed} (left), we show the distribution of systematically generated sequences (in \%) with their distance to opposed, depending on $m:=\sum_{i=1}^{n}a_i.$ We compare this scenario with the same setting for randomly generated sequences, shown in Figure~\ref{fig:SystematicDifferenceToOpposed} (right).

\emph{Observations:}
Systematically generated sequences have a slightly larger range of the ``distance to opposed'' than randomly generated sequences.
Moreover, when we generate dag sequences systematically, we obtain a 
significantly larger fraction of instances with a larger distance to opposed than for randomly generated sequences, and this phenomenon can be observed for all $m$.



\paragraph{Question 2: Do non-lexmax sequences possess a preference for large opposed distances?}

Since opposed sequences are easily solvable~\cite{BergerMueller-Hannemann-FCT2011}, we conjecture that sequences with a small distance to opposed might be easier solvable by the lexmax strategy than those with a large distance to opposed. If this conjecture were true, it would give us together with our findings from Question 1 one possible explanation for the observation that the randomly generated sequences have a larger fraction of efficiently solvable sequences by the lexmax strategy.

\emph{Observations:}
A separate analysis of non-lexmax sequences (that is, the subset of unsolved instances by the lexmax strategy), displayed in Figure~\ref{fig:OpposedDistanceNonLexMax}, gives a clear picture: yes! 
For systematically generated sequences with $n=9$, we observe in particular for instances with a middle density that the fraction of non-lexmax sequences becomes maximal for a relatively large distance to opposed.



\paragraph{Question 3: Can we solve real-world instances by the lexmax strategy?}

We consider real-world instances from different domains.
\begin{enumerate}
\item[a):] Ordered binary decision diagrams (OBDDs):
In such networks the outdegree is two, that is constant. 
This immediately implies that the corresponding sequences are opposed 
sequences, and hence can provably be solved by the lexmax strategy.
\item[b):] Food Webs: Such networks are almost hierarchical and therefore have a strong tendency to be acyclic
(``larger animals eat smaller animals''). In our experiments we analyzed 
food webs from the Pajek network library~\cite{Pajek-database04}.  
\item[c):] Train timetable network: We use timetable data of German Railways
from 2011 and form a time-expanded network. Its vertices correspond to departure and arrival events of trains, a departure vertex is connected by an arc with the arrival event corresponding to the very next train stop. Moreover, arrival and departure events at the same station are connected  whenever a transfer between trains is possible or if the two events correspond to the very same train.

\item[d):] Flight timetable network: We use the European flight schedule of 2010 and form a time-expanded network as in c).
\end{enumerate}

The characteristics of our real-world networks b) - d) are summarized in Table~\ref{tab:real-world}. The \emph{dag density} $\rho$ of a network is defined as $ \rho = m / { n \choose 2}$. To compare the distance to opposed for instances of different sizes, we normalize this value by the theoretical maximum ${ b \choose 2}$, where $b$ denotes the number of stream tuples, and so obtain a \emph{normalized distance to opposed}.
Without any exception, all real-world instances have been realized by the lexmax strategy.\\[-1ex]

\begin{table}[t]
\begin{center}
\begin{tabular}{l|r|r|r|c|c}
\phantom{xxxxxx} name and & & & & dag & norm. dist.\\
\phantom{xxx} kind of network & \phantom{xxx} $n$ \phantom{xxx} & \phantom{xx} $m$ \phantom{xxx}& \phantom{xx} $b$ \phantom{xx}&
\phantom{x}  density $\rho$ \phantom{x} & \phantom{x} to opposed\\\hline
burgess shale (b) & 142 \phantom{x} & 770 \phantom{x} & \phantom{x} 101 \phantom{x} & 0.08 \phantom{0.} & 0.40\\
chengjiang shale (b) & 85 \phantom{x}& 559 \phantom{x} & \phantom{x} 54 \phantom{x} & 0.16 \phantom{0.}& 0.50\\
florida bay dry (b) & 128 \phantom{x} & 2137 \phantom{x} & \phantom{x} 125 \phantom{x} & 0.26 \phantom{0.}& 0.32\\
cyprus dry  (b) & 71 \phantom{x} & 640 \phantom{x} & \phantom{x} 68 \phantom{x} & 0.26 \phantom{0.}& 0.43\\
maspalomas (b) & 24 \phantom{x} & 82 \phantom{x}& \phantom{x} 21 \phantom{x} & 0.30 \phantom{0.}& 0.30\\
rhode river (b) & 20 \phantom{x} & 53 \phantom{x}&  \phantom{x} 17 \phantom{x} & 0.28 \phantom{0.}& 0.42\\
train schedule 2011 (c) & 19359 \phantom{x} &  77201 \phantom{x} & \phantom{x} 18907 \phantom{x} & 0.0004 & 0.00\\
flight schedule 2010 (d) & 37800 \phantom{x}  &\phantom{x} 1324556 \phantom{x} & \phantom{x} 32905 \phantom{x} & 0.0019 & 0.00\\
\end{tabular}

\vspace*{1ex} 

\caption{\label{tab:real-world} Characteristics of our real-world test instances.}
\end{center}
\end{table}





\noindent
{\bf Back to theory.} Inspired by our observations in the systematic experiments,
we reconsidered forest sequences.
We can show that an arbitrary choice of a tuple in $V'_{min}$ in line 3 of Algorithm~\ref{alg:dag realization} solves the problem for forest sequences. 
  
\begin{theorem}[Realization of forest dags]\label{Th:Realization of forest dags}
 Let $S:={a_1 \choose b_1},\dots,{a_n \choose b_n}$ with $\sum_{i=1}^{n}a_i\leq n-1$ be a canonically sorted sequence containing $k>0$ source tuples. Furthermore, we assume that $S$ is not a source-sink-sequence. Consider an arbitrary stream tuple ${a_i \choose b_i}$ with $a_i\leq k.$ 
 $S$ is a dag sequence if and only if 
 \small 
 $$S':={0 \choose b_1-1},\dots,{0 \choose b_{a_{i}}-1},{0 \choose b_{a_{i}+1}},\dots,{0 \choose b_k},\dots,{a_{i-1} \choose b_{i-1}},{0 \choose b_{i}},{a_{i+1} \choose b_{i+1}},\dots,{a_n \choose b_n}$$
 \normalsize
 is a dag sequence.
 \end{theorem} 
 
Note, that sequence $S'$ may contain zero tuples. In this case, we delete these tuples and renumber the tuples from this new sequence $S':={a'_1 \choose b'_1},\dots,{a'_{n'} \choose b'_{n'}}$ from $1$ to $n'.$ Clearly, we have $\sum_{i=1}^{n'}a'_i \leq n-a_i-1\leq n'-1,$ because we deleted exactly the indegree of tuple ${a_i \choose b_i}$ in $S$ and  
 it is only possible to delete at most $a_i$ new zero tuples in $S'.$ Hence, Theorem \ref{Th:Realization of forest dags} results in a recursive algorithm. At each step, one has to choose an arbitrary stream tuple ${a_i \choose b_i}$ with indegree of at most $k$ and then to reduce $a_i$ largest sources by one and to set the indegree $a_i$ of this tuple to zero. On the other hand, the set $V_{min}$ of Theorem \ref{RealisationDagSequenzen} is a subset of the allowed tuple set in Theorem~\ref{Th:Realization of forest dags}. Hence, we get the following corollary. 
 
 \begin{corollary}[arbitrary tuple choice in $V_{min}$ for forest sequences]\label{co:sparse dag sequences}
 Let $S:={a_1 \choose b_1},\dots,{a_n \choose b_n}$ with $\sum_{i=1}^{n}a_i\leq n-1$ be a canonically sorted sequence containing $k>0$ source tuples. Furthermore, let $S'$ be defined as in Theorem \ref{RealisationDagSequenzen} where ${a_{i_{\ell}}\choose b_{i_{\ell}}}$ is an arbitrary tuple in $V_{min}.$\\
 $S$ is a dag sequence if and only if $S'$ is a dag sequence.
 \end{corollary}

\section{Randomized Algorithms}\label{RandomisierteAlgorithmen}

\subsection{Four versions of randomized algorithms}\label{fourRandAlgorithms}

The main idea for developing a randomized algorithm is the following.  In each trial use a randomly chosen topological sorting (a random permutation of the tuples) for a given sequence and then apply the linear-time realization algorithm as described in Section~\ref{Introduction} and justified by Lemma~\ref{TH:topologicalRealization}.
Clearly, it is not necessary to permute all tuples in a sequence. Instead we use a canonically sorted sequence and permute only the stream tuples. We denote this first naive version of a randomized algorithm by \emph{stream tuple permutation algorithm} (Rand I). A random permutation of a sequence of length $n$ can be chosen in $O(n)$ time, see for example \cite{Durstenfeld:1964}. Hence, one trial of the stream tuple permutation algorithm requires $O(m+n)$ time. 
This algorithm performs poorly since there are sequences with only a single realization among $(n-2)!$ many permutations of $n-2$ stream tuples. 
On the other hand, it is possible to restrict the number of possible topological sortings by the following lemma. 

\begin{lemma}[necessary criterion for the realizability of dag sequences]\label{Korollar:notwendiges Kriterium fuer Realisierbarkeit}
Let $S$ be a dag sequence. Denote the number of source tuples in $S$ by $q$ and the number of sink tuples by $s.$ Then it follows $a_i\leq \min\{n-s,i-1\}$ and $b_i\leq \min\{n-q,n-i\}$ for all $i\in \mathbb{N}_n$ for each labeling of $S$ corresponding to a topological order. 
\end{lemma}

\begin{figure}[h]
\centerline{\includegraphics[height=4cm]{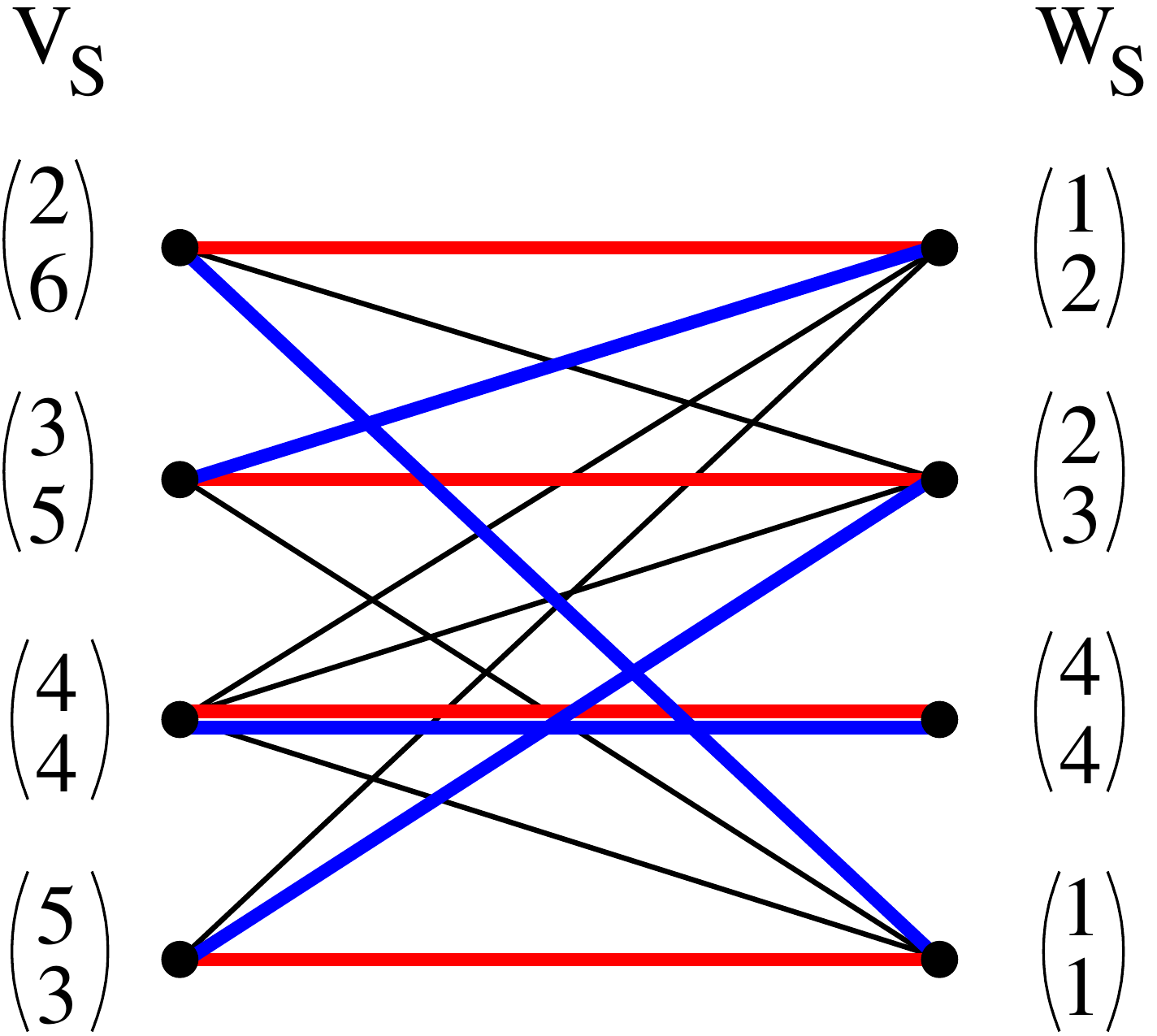}}
\caption{\label{fig:bounding-graph}Bounding graph $G_S$ for sequence $S:= { 0 \choose 3}, {0 \choose 1}, {1 \choose 2}, {2 \choose 3}, {4 \choose 4}, {1 \choose 1}, {1 \choose 0}, {2 \choose 0}, { 3 \choose 0}$. One perfect matching (thick red edges) leads to the topological order ${1 \choose 2}, {2 \choose 3}, {4 \choose 4}, {1 \choose 1}$ which is realizable, whereas another perfect matching (thick blue edges) gives the topological order ${1 \choose 1}, {1 \choose 2}, {4 \choose 4}, {2 \choose 3}$ which is not realizable.}
\end{figure}

Hence, a stream tuple ${a_i \choose b_i}$ can only be at position $j$ in a topological ordering if $a_j\leq \min\{n-s,i-1\}$ and $b_j\leq \min\{n-q,n-i\}$ is fulfilled. We define a bipartite \emph{bounding graph} $B_S=(V_S\cup W_S,E_S)$ for a given canonically sorted sequence as follows. 
We define $|S|-q-s$ vertices $v_i\in V_S$ with $i \in \{q+1,\dots,n-s\}$ where each vertex $v_i$ corresponds to an ``upper bound tuple'' ${\min\{n-s,i-1\}\choose \min\{n-q,n-i\}}$ for a stream tuple in $S$. Furthermore, we define $|S|-q-s$ vertices $w_i$ with $i \in \{q+1,\dots,n-s\}$ each corresponding to a stream tuple ${a_i \choose b_i}.$ The edge set $E_S$ is built as follows. Two vertices $v_i$ and $w_j$ are adjacent if and only if we find for ${a_j \choose b_j}$ that $a_j\leq \min\{n-s,i-1\}$ and $b_j\leq \min\{n-q,n-i\}.$ We show an example of the bounding graph (Figure~\ref{fig:bounding-graph}).

A perfect matching in this bounding graph gives us a possible topological sorting with respect to Lemma~\ref{Korollar:notwendiges Kriterium fuer Realisierbarkeit}. This means, we assign to each stream tuple ${a_j \choose b_j}$ in $S$ the number $i$ if and only if $(v_i,w_j)$ is a matching edge in the chosen perfect matching. Clearly, there does not exist a dag realization of sequence $S$ if $B_S$ does not contain a perfect matching. Unfortunately, the computation of the number of perfect matchings in a bipartite graph is known to be $\sharp P$-hard \cite{Valiant79}. 
On the other hand, there exists a polynomial-time algorithm for the problem of uniform sampling a perfect matching within a bipartite graph by Jerrum, Sinclair and Vigoda~\cite{JerrumSinclairVigoda04}. They use a Markov chain based algorithm. The number of necessary steps in this algorithm is measured by the so-called \emph{mixing time} $\tau_{\epsilon},$ where $\epsilon$ denotes the variation distance to the uniform distribution. They proved a worst case mixing time of $O(n^8(n \log{n} +\log{\frac{1}{\epsilon}})\log{\frac{1}{\epsilon}}).$ Up to know, we do not know if we really need a uniform distribution, but we do not want to eliminate certain topological orderings. 
Our second version of a randomized algorithm -- the \emph{bounding permutation algorithm} (Rand II) -- chooses in each trial a topological sorting by uniform sampling a perfect matching in $B_S$ and then applies the realization algorithm for a given topological order (Lemma~\ref{TH:topologicalRealization}).
For our experiments with very small instances, we sampled uniformly 
by enumerating all permutations of stream tuples. 

Our third randomized algorithm -- the \emph{opposed permutation algorithm} (Rand III) -- exploits the non-trivial result in Corollary~\ref{KorollarTopologischeRealisation} about opposed topological sortings. 
It uses for one trial, 
Algorithm~\ref{alg:dag realization} with a change in line $3$. 
We replace line $3$ by: ``\textbf{Sample a $v_j \in V'_{min}$ uniformly at random.}'' If possible, we restrict the set of $V'_{min}$ before line $3,$ i.e., we check for the largest $v_i\in V'_{min}$ whether the bounds of Lemma~\ref{Korollar:notwendiges Kriterium fuer Realisierbarkeit} are respected for later positions. Let $k$ denote the number of recursive calls up to the current one. 
Expressed in terms of the original sequence, 
we have to choose the $(q+k)$--th tuple in the topological sorting in the current iteration. 
If  $b_i= n-(q+k)$ for the lexicographical largest tuple ${a_i \choose b_i} \in V'_{min},$ then we set $V'_{min}:=\{ { a_i \choose b_i}\}.$ The reason is that a larger position is not possible at all for this tuple, because the upper bound for $b_i$ decreases strictly, as shown in Lemma~\ref{Korollar:notwendiges Kriterium fuer Realisierbarkeit}. 
At first glance it is not clear whether the restriction to a subset of permutations within the randomized algorithm really increases the chance to
draw a realizable topological sorting. 
This version of the algorithm only constructs dag realizations which possess an opposed topological sorting.
Hence, we also exclude possible topological sortings which are not opposed topological sortings. However, empirically this idea pays off.

Our fourth randomized version combines the opposed permutation algorithm with several \emph{reduction rules} which exploit the symmetric roles of in- and outdegrees and degree dominance of tuples. 
The following reduction rules can be used to simplify a given sequence.
Additional (similar) rules are possible, but we restrict our description to those rules which have been implemented and used in our experiments.

\begin{enumerate}
\item
\emph{Exploit symmetric roles of in- and outdegrees.}
If $|V'_{min}| = 1$, the reduction step 
in Algorithm~\ref{alg:dag realization} is safe (for any realizable sequence).
Since the problem is symmetric with respect to in- and outdegrees,
we can exchange their roles. This suggests to check the size 
of $V'_{min}$ from ``both sides''. If either of these sets has size one,
the corresponding reduction step is safe and should be preferably applied.

\item
\emph{Degree dominance of some tuple.}
Suppose that some $b_i$ is so large that this number matches the number of available stream and sink tuples, then vertex $v_i$ has to be connected with all current non-sources. Hence, sequence $S$ can be reduced by deleting a source tuple ${ 0 \choose b_i}$ or by updating a stream tuple ${ a_i \choose b_i}$ to 
a new sink ${ a_i \choose 0}$, respectively, and by subtracting one from all $a_j > 0$ with $i \neq j$. The symmetric reduction rule can be stated for a dominating $a_i$-value.



\item \emph{Dominating total degree of some stream tuple.}
Suppose there is a stream tuple with $a_i+b_i = n-1$.
Then we can conclude that this tuple has to be connected 
with all other tuples. It is unclear which stream tuples come before and 
which after ${a_i \choose b_i}$ in some realization. However, we can be sure
that it is connected with {\bf all} sources and all sinks 
(in particular $a_i \leq q$ and $b_i \leq s$ must hold).   
In order to ensure that later recursive reduction steps 
do not introduce parallel arcs, we only apply a more conservative reduction.
Namely, we connect the vertex $v_i$ only with sources and sinks for which
$a_i = 1$ or $b_i=1$, respectively. 
\end{enumerate}

%
We additionally apply these rules whenever applicable and call the randomized algorithm \emph{opposed permutation algorithm with reduction rules} (Rand IV). 

\subsection{Experimental comparison of randomized algorithms}





\paragraph{Experiment 1: Which randomized algorithm possesses the best success probability for one trial?}

\begin{figure}[t]
\vspace*{-4ex}
\begin{minipage}{.45\textwidth}
 \centering
 \includegraphics[width=1.2\textwidth]{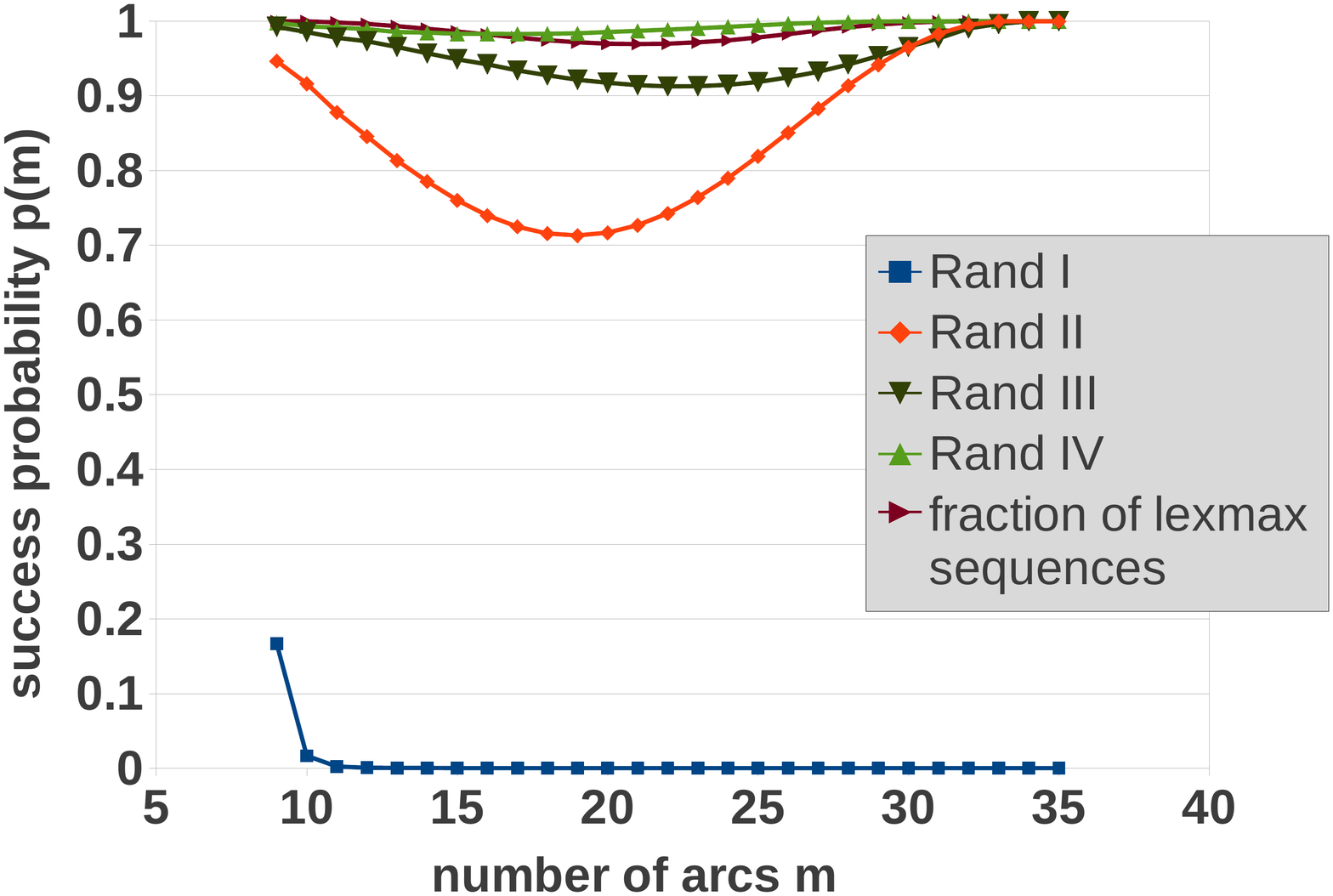}
 \caption{\label{fig:RandomizedSuccessProbability} Success probability $p(m)$ for all non-trivial sequences with $9$ tuples with four versions of randomized algorithms 
and the fraction of lexmax sequences.}
 \end{minipage}
\hfill
\begin{minipage}{.45\textwidth}
 \includegraphics[width=1.1\textwidth]{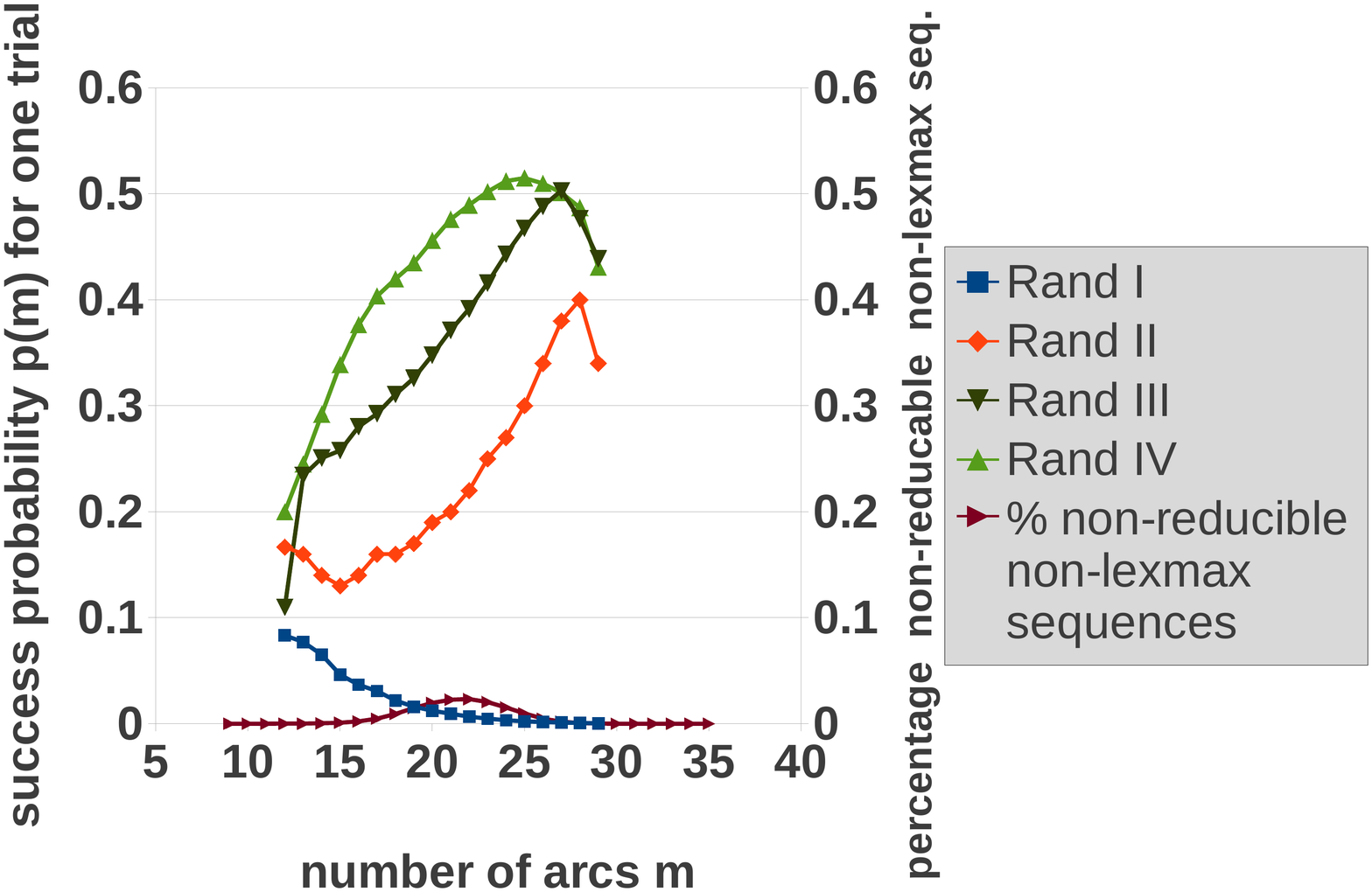}
 \caption{\label{fig:SuccessProbNonlexmaxRed} Success probability $p(m)$ for all non-reducible non-lexmax sequences of $9$ tuples with four versions of randomized algorithms and 
the percentage of non-reducible non-lexmax sequences in the set of all non-trivial sequences.}
\end{minipage}
 \end{figure}

We define the \emph{success probability} $p(m)$ as the probability that a given sequence $S:={a_1 \choose b_1},\dots,{a_n \choose b_n}$ with $m:=\sum_{i=1}^{n}a_i$ can be realized by a specified randomized algorithm in one single trial. In this experiment we test the four versions of our randomized algorithms with all non-trivial sequences (as defined in  Section~\ref{StoryExperiments}) of $9$ tuples, see Figure~\ref{fig:RandomizedSuccessProbability}. Moreover, we display the fraction of lexmax sequences to compare the deterministic lexmax strategy with our randomized strategy.\\[-1ex] 

\noindent
\emph{Observations:} Randomized version 4 (opposed permutation algorithm with reduction rules) clearly outperforms all other strategies.
We also observe that the success probability $p$ depends on the density
$m$ of the dag realizations. Sparse and dense dags have the best success probability. The deterministic lexmax strategy has almost the same success probability as our best randomized version. Of course, we can repeat a randomized algorithm and thereby boost the success rate which is not possible for the deterministic variant. Nevertheless the good performance of the simple lexmax strategy is quite remarkable, it clearly outperforms an arbitrary strategy to choose in line 3 of Algorithm~\ref{alg:dag realization} an element from $V'_{min}$ (realized in randomized version 3).

\paragraph{Experiment 2: We consider the success probability for all randomized algorithms in the case of non-lexmax sequences which are not reducible by our reduction rules.}

Noting that an impressively large fraction of sequences is efficiently solvable by the deterministic lexmax strategy combined with our reduction rules,   
we should ask: How well do our randomized algorithms perform for the remaining difficult cases, that is for \emph{non-reducible non-lexmax sequences}? Actually, this is indeed the most interesting question, because the best approach for realizing a given sequence $S$ would be: first to test, whether $S$ is a reducible lexmax sequence. Only if this is not the case, one would take a randomized algorithm. Hence, we now determine the success probability $p(m)$ for all non-reducible non-lexmax sequences, see Figure~\ref{fig:SuccessProbNonlexmaxRed}.\\[1ex]

\noindent
\emph{Observations:} As in the previous experiment, randomized version 4 has the overall best success probability $p$, but in sharp contrast we observe a completely different dependence on $m$. One possible explanation could be that for high densities our reduction rules have been applied more often.
Note that the overall percentage of non-reducible non-lexmax sequences in the set of all non-trivial sequences with 9 tuples is so tiny (see the brown curve in Figure~\ref{fig:SuccessProbNonlexmaxRed}) --- in particular for low densities --- that we can realize after two or three trials almost all sequences.

\section{Conclusion}\label{Conclusion}

In this paper we have studied the performance of a simple linear-time heuristic to solve the NP-complete dag realization problem and several randomized variants. 
We give a brief summary of our main observations.

 \begin{enumerate}
 \item Dag sequences $S$ with sparse or dense densities are almost always 
       lexmax sequences.
 \item Dag sequences with a small distance to opposed $d(S)$ are to a large extent  lexmax sequences.
 \item There is a good chance to realize a dag sequence by the lexmax strategy, especially for acyclic real-world networks.
 \end{enumerate}

For a given (real-world) sequence we propose the following recipe: Choose Algorithm~\ref{alg:dag realization} with lexmax strategy and apply the reduction rules 1-3. If this run is unsuccessful apply version 4 of our randomized algorithms, i.e.\ the opposed permutation algorithm with reduction rules.
 For most dag sequences in practice this will give us a pretty fair chance to find a realization.
The surprisingly broad success of the lexmax strategy suggests that
there might be further subclasses of instances where it runs provably correct.
In future work we would like to characterize the class of instances for which the lexmax strategy works provably correct.


\providecommand{\bysame}{\leavevmode\hbox to3em{\hrulefill}\thinspace}
\providecommand{\MR}{\relax\ifhmode\unskip\space\fi MR }
\providecommand{\MRhref}[2]{%
  \href{http://www.ams.org/mathscinet-getitem?mr=#1}{#2}
}
\providecommand{\href}[2]{#2}

\newpage

\centerline{\Huge \bf Appendix}

\begin{appendix}
\section{Proofs}

In this section we present the proofs for our theoretical results.
For the first lemma, recall the corresponding setting from 
Section~\ref{Introduction}. 

Let $S:={a_1 \choose b_1},\dots,{a_{q+1} \choose b_{q+1}},\dots,{a_n \choose b_n}$ be an arbitrary sequence with a given topological order.
Without loss of generality, we may assume that the source tuples come first in the prescribed numbering and are ordered decreasingly with respect to their $b_i$ values. 
Let $S':={a_1 \choose b_1-1},\dots,{a_{a_{q+1}} \choose b_{a_{q+1}}-1},\dots,{a_q \choose b_q},{0 \choose b_{q+1}},\dots,{a_n \choose b_n}.$ If we get zero tuples in $S',$ then we delete them and denote the new sequence for simplicity also by $S'.$ Furthermore, we label this sequence with a new numbering starting from one to its length and consider this sorting as the given topological ordering of $S'$.\\[-1ex]

\noindent
{\bf Lemma~\ref{TH:topologicalRealization}.}
\emph{$S$ is a dag sequence for a given topological order $\Leftrightarrow$ $S'$ is a dag sequence for its corresponding topological order.
}

\begin{proof}
$\Leftarrow:$ Trivial.\\
$\Rightarrow:$ We consider a dag realization for the given topological ordering of dag sequence $S.$ Clearly, we find at least $a_{q+1}$ sources. This is true, because a first vertex with non-empty incoming neighborhood set in a topological sorting (a sink or a stream vertex) of a dag can only possess sources in its incoming neighborhood set. Otherwise, this numbering is not a topological sorting. Assume, there is no dag realization for the given topological order, such that the $a_{q+1}$ largest sources are connected with vertex $v_{q+1}.$ In this case, we consider a dag realization $G$ to this topological order such that the maximum possible number of largest sources is connected with vertex $v_{q+1}.$ Then we have two sources $v_{i}$ and $v_{j}$ with $(v_i,v_{q+1})\notin A,$ $(v_j,v_{q+1}\in A),$ $b_i > b_j$ and $i,j<q+1.$  Since $b_i>0$ and $v_{q+1}$ is the first non-source tuple, there is a non-source vertex $v_k$ $(k>q+1)$ with $(v_i,v_k)\in A$ and $(v_j,v_k)\notin A.$ We define a new digraph $G^{*}:=(G\setminus \{v_i,v_k\}\cup \{v_j,v_{q+1}\}) \cup (\{v_i,v_{q+1}\}\cup \{v_j,v_k\}).$ Obviously, $G^{*}$ is a dag realization for the given topological order of sequence $S.$ Contradiction to the assumption that $G$ is a dag realization with the maximum possible number of largest sources for vertex $v_{q+1}$. Hence, there exists a dag realization $G$ to the given topological order such that vertex $v_{q+1}$ has in its incoming neighborhood set only the $a_{q+1}$ largest sources from the set of all sources $v_i$ with $i<q+1.$ We delete the incoming neighborhood set of vertex $v_{q+1}$ and yield a dag realization for sequence $S'$ for its given topological ordering. \hfill$\Box$
\end{proof} 

The existence of a simple solution for forest sequences is not so surprising as there is a simple approach to construct a dag realization if there is one. First, one can apply a digraph realization algorithm. When we do not find a digraph realization, there also cannot be a dag realization. Assume, we have a digraph realization $G=(V,A).$ If $G$ possesses no directed cycle then it is a dag realization and we are ready. Let us assume, $G$ has at least one directed cycle. In this case, there exist at least two weak components (i.e.\ connected components in the underlying undirected graph), because the underlying undirected graph is not a forest. Hence, we can choose an arc $(v_1,v_2)$ of the directed cycle in the first component and a further arc $(v_3,v_4)$ from the second weak component. We construct the new digraph $G':=(V,A')$ with $A':=A\setminus \{(v_1,v_2),(v_3,v_4)\}\cup \{(v_1,v_4),(v_3,v_2)\}.$ We apply a sequence of such steps (at most $n$ steps) until we get an acyclic dag realization. This is possible because as long as we can find a directed cycle we also have more than one weak component.
(Note, that the underlying graph is not necessarily a simple graph. It can contain parallel edges corresponding to directed $2$-cycles of the initial dag realization $G$.) Hence, we can conclude that each forest sequence which is a digraph sequence is also a dag sequence. Clearly, this can be decided in polynomial time.\\[-1ex]

Note that sequence $S'$ may contain zero tuples. In this case, we delete these tuples and renumber the tuples from this new sequence $S':={a'_1 \choose b'_1},\dots,{a'_{n'} \choose b'_{n'}}$ from $1$ to $n'.$ Clearly, we have $\sum_{i=1}^{n'}a'_i \leq n-a_i-1\leq n'-1,$ because we deleted exactly the indegree of tuple ${a_i \choose b_i}$ in $S$ and  
 it is only possible to delete at most $a_i$ new zero tuples in $S'.$ Hence, Theorem \ref{Th:Realization of forest dags} results in a recursive algorithm. At each step, one has to choose an arbitrary stream tuple ${a_i \choose b_i}$ with indegree of at most $k$ and then to reduce $a_i$ largest sources by one and to set the indegree $a_i$ of this tuple to zero. On the other hand, the set $V_{min}$ of Theorem \ref{RealisationDagSequenzen} is a subset of the allowed tuple set in Theorem~\ref{Th:Realization of forest dags}. Hence, we get Corollary~\ref{co:sparse dag sequences}.\\

 \noindent
 {\bf Theorem~\ref{Th:Realization of forest dags} (Realization of forest dags).} 
 \emph{ Let $S:={a_1 \choose b_1},\dots,{a_n \choose b_n}$ with $\sum_{i=1}^{n}a_i\leq n-1$ be a canonically sorted sequence containing $k>0$ source tuples. Furthermore, we assume that $S$ is not a source-sink-sequence. Consider an arbitrary stream tuple ${a_i \choose b_i}$ with $a_i\leq k.$\\
  $S$ is a dag sequence if and only if 
  \small 
  $$S':={0 \choose b_1-1},\dots,{0 \choose b_{a_{i}}-1},{0 \choose b_{a_{i}+1}},\dots,{0 \choose b_k},\dots,{a_{i-1} \choose b_{i-1}},{0 \choose b_{i}},{a_{i+1} \choose b_{i+1}},\dots,{a_n \choose b_n}$$
  \normalsize
  is a dag sequence.
 }
  
 \begin{proof} (of Realization of forest dags)
 $\Leftarrow:$ Trivial.\\
 $\Rightarrow:$ Let $S$ be a dag sequence with $k\geq 1$ source tuples. We consider a dag realization $G$ such that we have a minimum number of weak components. Clearly, the underlying undirected graph is then a forest without undirected cycles. Furthermore, we consider a dag realization $G$ as described where the incoming neighborhood set of vertex $v_i$ consists of a maximum possible number of sources. Assume, there is a vertex $v_{i^{-}}\in N_G^{-}(v_i)$ which is not a source. 
(The notation $N_G^{-}(v)$ describes the in-neighborhood of vertex $v$ in $G$.)
Then we can conclude that there exists a source $q$ and a vertex $v_j$ with $(q,v_j)\in A$ but $(q,v_i)\notin A,$ because we have $d_{G}^{-}(v_i)=a_i\leq k$ by our assumption. We distinguish between two cases.\\[-1ex]
 
\noindent 
{\bf case 1:} There exists no underlying undirected path between vertices $v_j$ and $v_i.$\\
{\bf case 2:} There exists exactly one underlying undirected path $P$ between vertices $v_j$ and $v_i.$\\
 
 Note that there cannot be more than one underlying undirected path, because the underlying graph of $G$ is by our assumption a forest. Let us start with case $1.$ There cannot be an underlying undirected path between vertices $q$ and $v_i,$ otherwise we would find the excluded undirected path between $v_i$ and $v_j,$ because $q$ is adjacent to $v_j.$
 We construct the dag $G'=(V,A')$ with $A':=A\setminus \{(q,v_j),(v_{i^{-}},v_i)\}\cup \{(q,v_i),(v_{i^{-}},v_j)\}.$ Consider Figure \ref{fig:ForestRealizationCase1A}. Digraph $G$ is indeed a dag, because $G$ does not contain an underlying undirected path between $v_i$ and $q$ and not between $v_{i^{-}}$ and $v_j$ by our assumptions. Hence, we did not construct an underlying undirected cycle and clearly no directed cycle.  
 
 \begin{figure}
 \centering
 \includegraphics[width=6cm]{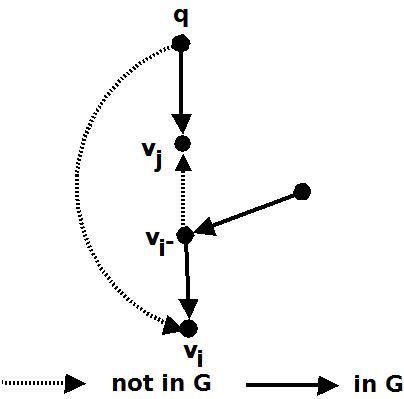}
 \caption{\label{fig:ForestRealizationCase1A} Case $1:$ no underlying undirected path between $q$ and $v_i$ in $G.$} 
 \end{figure} 
 
 But then $G'$ is a dag realization with a minimum number of weak components and a larger number of sources in the neighborhood set of $v_i$ than in dag $G.$ Contradiction!
 It remains to consider case $2.$ Since vertex $v_i$ is a stream tuple we define the following dag $G'=(V,A')$ with $A':=A\setminus \{(q,v_{j}),(v_{i^{-}},v_i),(v_i,v_{i^{+}})\}\cup \{(q,v_i),(v_{i^{-}},v_{i^{+}}),(v_{i},v_{j})\}$ as can be seen in Figure~\ref{fig:ForestRealizationCase2}.
 
 \begin{figure}[h]
 \centering
 \includegraphics[width=6cm]{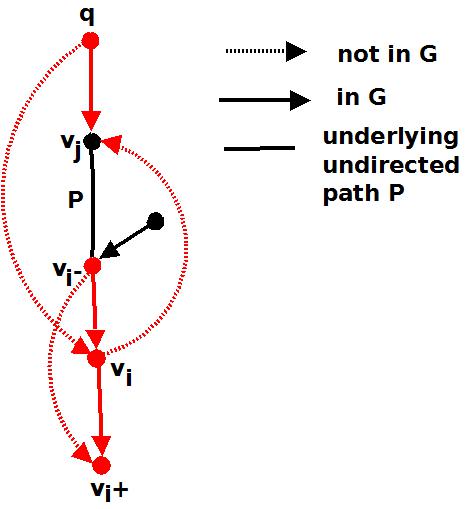}
 \caption{\label{fig:ForestRealizationCase2} Case $2:$ one unique underlying undirected path $P$ between $q$ and $v_i.$}
 \end{figure} 
 
 Note, that $v_j$ and $v_{i^{-}}$ are not necessarily distinct vertices. If this is the case case, then we replace in $A'$ vertex $v_{i^{-}}$ by $v_j.$ Since we destroyed by our construction all underlying unique paths in $G$ between $v_i$ and $q$, between $v_{i^{+}}$ and $v_{i^{-}}$ and between $v_{j}$ and $v_{i},$ digraph $G'$ is indeed acyclic and possesses a minimum number of weak components. On the other hand vertex $v_i$ is connected with a larger number of sources as in $G.$ Contradiction!\\
 
 Hence, we can assume that there exists a dag realization $G=(V,A)$ with a minimum number of weak components such that the incoming neighborhood set of vertex $v_i$ only contains sources. We consider a dag realization such that vertex $v_i$ is connected with the maximum possible number of largest sources. Assume, there is a source $q'>q$ such that $(q,v_i)\in A$ and $(q',v_i)\notin A.$ Then there exists a further vertex $v_j$ with $(q',v_j)\in A.$ We distinguish again between two cases. If there does not exist an underlying undirected path $P$ between $q$ and $v_j$ or between $q'$ and $v_i,$ we define the new dag $G':=(V,A')$ with $A':=A\setminus \{(q,v_i),(q',v_j)\}\cup \{(q',v_i),(q,v_j)\}$  with a minimum number of weak components but with one larger source connected with $v_i$ than in $G$ (see Figure~\ref{fig:largestSources}). Contradiction!
 
 \begin{figure}[h]
 \centering
 \includegraphics[width=11cm]{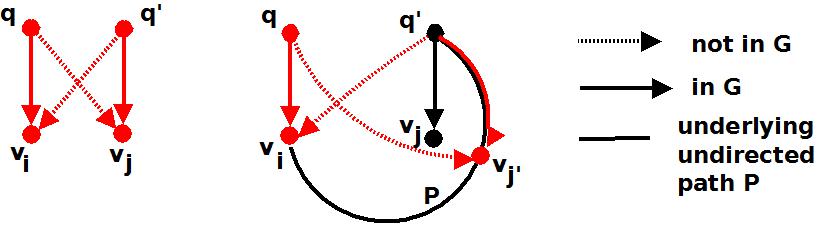}
 \caption{\label{fig:largestSources} A larger source $q'$ is not connected with vertex $v_i.$}
 \end{figure}
 
 Hence, we next assume that there is one underlying undirected path $P=q',v_{j^{'}},\dots,v_i$ between $q'$ and $v_i.$ (A further path between $q$ and $v_j$ cannot exist, because in this case we would find an underlying cycle.) Note, that it is possible that we find $v_{j^{'}}=v_j.$ In this case we replace in the following steps $v_{j^{'}}$ by $v_j.$ We define the new dag realization $G'=(V,A')$ with $A':=A\setminus \{(q,v_i),(q',v_{j^{'}})\}\cup \{(q',v_i),(q,v_{j^{'}})\}$ with a minimum number of weak components, because we destroyed by our construction the underlying unique paths from $q$ to $v_{j^{'}}$ and from $q'$ to $v_i.$ Dag $G'$ possesses one more of the largest sources connected to $v_i$ than $G.$ Contradiction! 
 As a last case it remains, that there could exist an underlying undirected path $P=q,\dots,v_j$ from $q$ to $v_j,$ see Figure \ref{fig:largestSourcesLastCase}.
 
 \begin{figure}
 \centering
 \includegraphics[width=8cm]{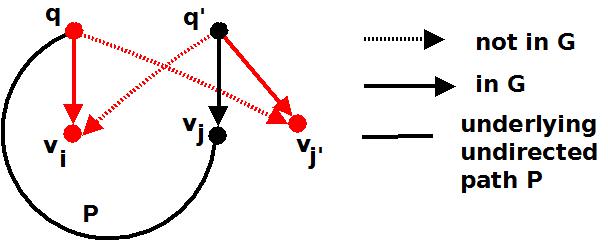}
 \caption{\label{fig:largestSourcesLastCase} A larger source $q'$ is not connected with vertex $v_i$ and there exists an underlying path $P$ between $q$ to $v_j.$}
 \end{figure}
 
 Since $q'$ is a larger source than $q,$ there exists a further vertex $v_{j^{'}}$ with $(q',v_{j^{'}})\in A.$ Then we construct the dag realization $G'=(V,A')$ with 
 $A':=A\setminus \{(q,v_i),(q',v_{j^{'}})\}\cup \{(q',v_i),(q,v_{j^{'}})\}.$ Indeed, we destroyed in $G$ the unique paths between $q$ and $v_{j^{'}}$ and between $q'$ and $v_i.$ Hence, $G'$ is a dag with a minimum number of weak components but with one more of the largest sources connected to $v_i$ than in $G.$ Contradiction!\\
 So, there exists a dag realization $G$ such that vertex $v_i$ has in its incoming neighborhood set only largest sources. We delete 
the arcs from these sources to $v_i$ in $G,$ and get a dag realization $G'$ with dag sequence $S'.$ \hfill$\Box$
 \end{proof}

\noindent
{\bf Lemma~\ref{Korollar:notwendiges Kriterium fuer Realisierbarkeit} (necessary criterion for the realizability of dag sequences). }
\emph{Let $S$ be a dag sequence. Denote the number of source tuples in $S$ by $q$ and the number of sink tuples by $s.$ Then it follows $a_i\leq \min\{n-s,i-1\}$ and $b_i\leq \min\{n-q,n-i\}$ for all $i\in \mathbb{N}_n$ for each labeling of $S$ corresponding to a topological order. 
}

\begin{proof}
Let $S:={a_1 \choose b_1},\dots,{a_n \choose b_n}$ be a labeling of $S$ corresponding to a topological sorting of a dag realization $G.$ Assume, there is a $j\in \mathbb{N}_n$ with $a_j>\min\{n-s,j-1\}.$ (Case $b_j>\min\{n-q,n-j\}$ can be done analogously.) $G$ is a subdigraph of a complete dag $G^*$ with topological sorting $v_1, \dots, v_n.$ Clearly, we have $d_{G^*}^-(v_j)=j-1.$ We distinguish between two cases. If we have $\min\{n-s,j-1\}=n-s,$ then it follows $a_j=d^-_{G}(v_j)>n-s.$ Then the incoming neighborhood set $N^-_{G}(v_j)$ consists of more than $n-s$ vertices -- in contradiction to the fact that $N^-_{G}(v_j)$ contains at most $n-s$ vertices. Let us now assume $\min\{n-s,j-1\}=j-1.$ Then we get $d^-_{G^{*}}(v_j)=j-1<a_j=d^-_{G}(v_j)$ -- a contradiction to our assumption that $G$ is a subdigraph of $G^*.$ \hfill$\Box$
\end{proof}

\section{Example for Algorithm~\ref{alg:dag realization}}
\label{app:examples}

\begin{figure}[H]
\centerline{\includegraphics[trim=3cm 2cm 1cm 2cm,clip,width=1.2\textwidth]{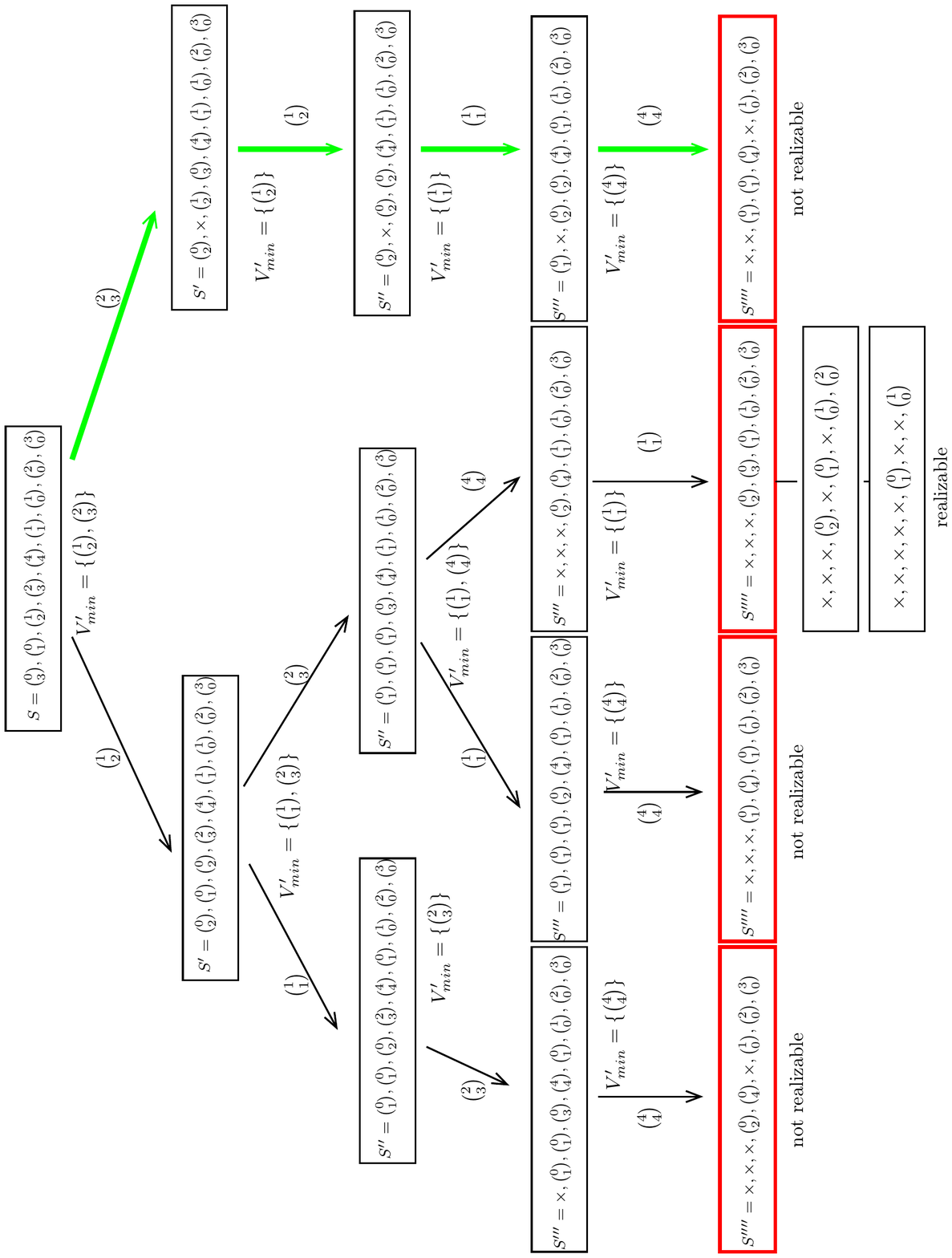}}
\caption{\label{fig:example} Recursion tree for Example~\ref{example}. The symbol $\times$ here denotes tuples of $S$ which have been deleted after 
being reduced to ${ 0 \choose 0}$. The forth tree level where the original sequence is reduced to a source-sink sequence is marked with red boxes.} 
\end{figure}

\emph{Example~\ref{example}.}
Consider the sequence $S = {0 \choose 3},{0 \choose 1},{1 \choose 2},{2 \choose 3},{4 \choose 4},{1 \choose 1}, {1 \choose 0}, {2 \choose 0}, {3 \choose 0}$.
Figure~\ref{fig:example} shows the recursion tree of 
Algorithm~\ref{alg:dag realization} for this instance. 
The symbol $\times$ here denotes tuples of $S$ which have been deleted after 
being reduced to ${ 0 \choose 0}$.
The rightmost path (green) corresponds to the lexmax strategy, not leading to a realization. 

\newpage

\section{Further 
Supporting Material}\label{app:material}

\begin{table}
\begin{center}
\begin{tabular}{c|r|r|r|r|}
    &     & \# non-trivial & \# non-lexmax & \# reduced non-lexmax\\  
$n$ & $m$ &    sequences   & sequences     & sequences\\\hline
9 &	9 &	1469 & 0 & 0 \\
9 &	10 &	4566 & 0 & 0 \\
9 &	11 &	12284 & 22 & 0 \\
9 &	12 &	29350 & 106 & 1 \\
9 &	13 &	63411 & 418 & 12 \\
9 &	14 &	124958 & 1255 & 54 \\
9 &	15 &	226343 & 3148 & 146 \\
9 &	16 &	379089 & 6759 & 337 \\
9 &	17 &	590302 & 12916 & 763 \\
9 &	18 &	855830 & 21825 & 1492 \\
9 &	19 &	1155082 & 32707 & 2394 \\
9 &	20 &	1451117 & 43519 & 3175 \\
9 &	21 &	1695124 & 51757 & 3673 \\
9 &	22 &	1839040 & 55112 & 3757 \\
9 &	23 &	1846761 & 52270 & 3300 \\
9 &	24 &	1710913 & 43800 & 2475 \\
9 &	25 &	1453602 & 31678 & 1549 \\
9 &	26 &	1124025 & 19399 & 754 \\
9 &	27 &	783283 & 9767 & 286 \\
9 &	28 &	485528 & 3917 & 89 \\
9 &	29 &	262909 & 1164 & 14 \\
9 &	30 &	121343 & 235 & 0 \\
9 &	31 &	46183 & 25 & 0 \\
9 &	32 &	13867 & 0 & 0 \\
9 &	33 &	3059 & 0 & 0 \\
9 &	34 &	448 & 0 & 0 \\
9 &	35 &	36 & 0 & 0 \\
\end{tabular}
\caption{\label{tab:row-data} Systematic experiments with $n=9$. For $m \in \{ 9, \dots, 35\}$, we show the number of non-trivial sequences (i.e. sequences which have at least one stream tuple, column 3), the number of sequences where the pure lexmax strategy fails (column 4), and finally the number of sequences where the lexmax strategy combined with our reduction rules fails (column 5).}
\end{center}
\end{table}

\end{appendix}

\end{document}